# Open Data Quality Evaluation: A Comparative Analysis of Open Data in Latvia

Anastasija Nikiforova ORCID: 0000-0002-0532-3488
Faculty of Computing, University of Latvia, 19 Raina Blvd., Riga, LV-1586, Latvia
Nikiforova.Anastasija@gmail.com

**Abstract**. Nowadays open data is entering the mainstream - it is free available for every stakeholder and is often used in business decision-making. It is important to be sure data is trustable and error-free as its quality problems can lead to huge losses. The research discusses how (open) data quality could be assessed. It also covers main points which should be considered developing a data quality management solution. One specific approach is applied to several Latvian open data sets. The research provides a step-by-step open data sets analysis guide and summarizes its results. It is also shown there could exist differences in data quality depending on data supplier (centralized and decentralized data releases) and, unfortunately, trustable data supplier cannot guarantee data quality problems absence. There are also underlined common data quality problems detected not only in Latvian open data but also in open data of 3 European countries.

## Introduction and Motivation

According to Siemens CEO Joe Kaeser, data is the 21st century's oil (WEB, a). Data quality issue is actual and widely discussed in scientific researches for many years. However, nowadays, data quality issue became even more popular as open data is entering the mainstream (WEB, f).

According to (Huijboom and Van den Broek, 2011), nowadays, governments around the world define and implement "open data strategies" to increase transparency, participation and/or government efficiency. The commonly accepted opinion (e.g. (Gruen et al., 2014), (Huijboom and Van den Broek, 2011), (Vickery, 2011), (Welle Donker and van Loenen, 2017)) is that the publishing of data in a reusable format can enable new services and yield innovative businesses, improve quality of services, reduce the cost of existing services, strengthen citizen engagement and/or government efficiency, engender greater trust in government increasing transparency (in case of open government data). However, there are barriers of open data policy implementation and one of the most important barriers achieving this goal is data quality issues (Janssen et al., 2012), (Craglia et al., 2010).

Open data is usually used with common assumption that it is of high quality and is ready to process without additional activities (such as quality checks). It is often used in analysis making important (business) decisions according to analysis results. Its quality influences decision-making and data quality problems can lead to huge losses. For instance, in 2018 Gartner research has found that organizations believe poor data quality to be responsible for an average of $15 million per year in losses. This fact was proved for many times (e.g. (Friedman and Judah, 2013), (Kelly, 2009), (Moore, 2018), (WEB, k)). Moreover, such tendency is observed for many years (even in 1992, more than 60 percent of the firms (among 500 medium-size corporations with annual sales of more than $20 million) had problems with data quality (Arnold, 1992)), and data quality state doesn't significantly improve through the last years.

"Data quality" concept has many definitions. However, data is generally considered of high quality if it fits for its intended uses (by consumers/customers) in operations, analytics, decision-making, and planning. It is important in both decisional and operational processes (Batini and Scannapieco, 2016). Data is fit for use if it is free of defects and possess desired features (Redman, 2001), (Wang and Strong, 1996).

It should be also mentioned, that data preparation for analysis (which includes data quality checking) is one of the most time-consuming and difficult activities, which should be taken before data could be used in analysis

(Pyle, 1999). According to Gabernet and Limburn (2017), 80 percent of a data scientist's valuable time is spent simply finding, cleansing, and organizing data, leaving only 20 percent to perform analysis. Moreover, according to TDQM, "1-10-100" rule is valid for data quality: one dollar spent on prevention will save 10 dollars on appraisal and 100 dollars on failure costs (Ross, 2017).

Nowadays, most of the sources, which deal with (open) data quality issue, focuses on (open) data quality characteristics informal definition and their values measuring but mechanisms for specifying data quality characteristics in formalized languages usually are not considered. In previous research (Bicevska et al., 2017) author's colleagues have already described the main points of suggested approach, which could solve data quality problems. This research provides short overview on this mechanism and demonstrates its application to open data checking its quality. All these facts approve importance of (open) data quality topic.

The first aim of this paper is to analyse Latvian open data quality, clarifying whether publicly available data sets are trustful.

The second aim is to apply in (Bicevska et al., 2017) proposed approach to real Latvian open data sets verifying its appropriateness to the specified task.

The third aim is to check Vetrò et al. (2016) assumption, according to which, there could exist differences between data quality of data provided by governments (the centralized data release, at national level) and municipalities (decentralized data release).

The paper deals with following issues: overview on the related researches (Section 2), problem statement (Section 3), a description of the proposed solution (Section 4) open data analysis (Section 5), results (Section 6), future plans (Section 7).

## State of the Art

Existing researches and practical solutions on data quality can be divided into several groups: (a) data quality researches defining data quality dimensions, (b) quality assessment of open data portals and/or Open Government Data (OGD), (c) data quality assessment frameworks, (d) quality assessment of linked data, (d) data quality guidelines.

One of the main theories in data quality area is Total Data Quality Management (TDQM) (for overview of existing methodologies see (Batini et al., 2009)). According to TDQM, data quality has many dimensions for data users and nowadays, assessment using data quality dimensions is widely used and most of the existing researches focuses on data quality dimensions and their application to data sets ((Färber et al., 2016), (Ferney et al., 2017), (Redman, 2001), (Wang and Strong, 1996)). Different researches define different quality dimensions and their groupings, for instance, in 2013 Data Management Association International UK Working Group defined 6 data quality dimensions such as completeness, uniqueness, timeliness, validity, accuracy, consistency (WEB, n). At the same time Redman (2001) recommends using 8 data quality dimensions: accuracy, completeness, redundancy, readability, accessibility, consistency, usefulness and trust (each dimension has list of criteria). It can be observed that some of these dimensions overlap while some of them are unique for specific research.

Ferney et al. (2017) proposes analysing data sets according to three dimensions - traceability, completeness, compliance. The measuring of the metrics is done with a software called RapidMiner, which allows to do data mining. RapidMiner supports the process of obtaining, processing, storing and assessing data obtained from the analysed open data portals. Analysis results provide users with general information about data sets quality focusing on metadata (if metadata is provided) and null values in rows and columns. Unfortunately, definition of specific requirements for specific fields and syntax checks aren't possible.

Färber et al. (2016) focuses on openly available knowledge graphs such as DBpedia, Freebase, OpenCyc, Wikidata, and YAGO. According to the authors, knowledge graphs have not been subject to an in-depth comparison so far, thus they provide data quality criteria, according to which knowledge graphs can be analysed. Authors use dimensions defined by Wang and Strong (1996), Bizer (2007) and Zaveri et al. (2016), applying them to the above-mentioned knowledge graphs. This research focuses on knowledge graphs and can't be customized and used to analyse separate data sets. Moreover, it requires deep knowledges about knowledge graphs.

Most of the existing solutions (frameworks) examine open data portals. For example, Neumaier (2015) and Umbrich et al. (2014) provide overview of automated quality assessment frameworks, which allows discovering and measuring quality and heterogeneity issues in open data portals.

According to (Neumaier, 2015), there neither exists a comprehensive and objective report about the actual state and quality of Open Data portals, nor is there a framework to continuously monitor the evolution of these portals. To solve this problem the authors of (Neumaier, 2015) and (Umbrich et al., 2014) define two objectives: (1) to discover, point out and measure quality and heterogeneity issues in data portals, (2) to develop automated quality assessment framework providing possibility to monitor and assess the evolution of quality metrics over time. They present various detected quality issues in open data in result of monitoring and assessing the quality of three data portals. The authors also define six quality metrics: retrievability, usage, completeness, accuracy, openness and contactability (as these dimensions doesn't correspond with widely used, they define every dimension from their perspective). They also introduce automated assessment framework that periodically monitors the content of CKAN portal and compute a set of quality metrics to gain insights about the evolution of the (meta-) data. They also propose a common mapping for metadata occurring on analysed portals software frameworks in order to improve the comparability and interoperability of portals running these different software frameworks. These solutions are useful examining the whole portal quality (but not several open data sets).

The Open Data Institute report (WEB, l) explores the Sustainability of Open Data Portals across Europe. The authors underline that the most of open data portals do not have coherent strategies for sustainability that address each aspect of how a portal functions: the governance, financing, architecture and operations that make a portal sustainable, as well as the metrics that can be used to monitor and evaluate progress. This report provides recommendations in each of these five key areas which are drawn from in-depth interviews with portal owners across Europe (seven countries).

Another example is the PoDQA project (Caro et al., 2007) - a data quality model for Web portals based on more than 30 data quality attributes. This solution is orientated towards evaluating data quality from the perspective of the data consumer. Its main functions are: calculating the data quality level in a given portal, providing the user with information about the data quality in a given portal. One of the main benefits is that this model can be used by portal users and their developers. Unfortunately, such high number of attributes can lead to misunderstandings as some criteria are similar and sometimes the difference between them is minimal (possibly, DAMA classification could be more appropriate) and this time users should choose appropriate dimensions, which would be used in data quality analysis.

These solutions also aren't appropriate to examine specific data sets as, for instance, (Caro et al., 2007) provides ranking of the portals according to data quality but it is impossible to get details about data quality issues in specific data set. Despite the fact proposed solutions are mainly automatized and doesn't require programmers or other IT specialists involving making analysis, it could be difficult to use them without appropriate knowledges of data quality dimensions, which should be chosen to be applied to data sets analysing data quality.

Some researches, for example, (Kučera et al., 2013), (Vetrò et al., 2016) and (WEB, l) define guidelines for data suppliers and open data portals, which should be considered before data is published. These researches mainly focus on Open Government Data (OGD). According to (Kučera et al., 2013) and (Vetrò et al., 2016), number of the OGD catalogues has been established over the past years, but recent experience shows that the quality of the catalogue records might affects the ability of users to locate the data of their interest. Vetrò et al. (2016) emphasizes there are no a comprehensive theoretical framework, and most of the evaluations focuses on open data platforms, rather than on datasets. Thus, authors set up a framework of indicators to measure the quality of Open Government Data using set of seven data quality dimensions. This set was created in result of survey of 15 developers. It can't be observed as the list of the most important (and appropriate) dimensions for this task, and there is no possibility to define additional quality checks or specify already defined according to stakeholder needs. The authors propose framework, which is based on SPDQM methodology, and demonstrate results of OGD data quality analysis. Based on analysis results, the authors provide guidelines to solve detected issues. This solution can be applied to specific data sets and it provides even graphical results but analysis supposed to be done involving data quality experts - users can specify quality dimensions, which should be included in analysis of data sets. Moreover, according to developers of this solution, it is not able to check correctness of specific data formats. It also isn't known if there are available protocols, which could be used detecting specific records, which must be revised and which quality must be improved. Similarly to (Neumaier, 2015) and (Umbrich et al., 2014), proposed solutions are useful improving data quality, but they should be used by portals holders and inspectors preparing data for publishing rather than data consumers checking data quality before it will be used in analysis.

Other researches mainly deal with linked (open) data quality - (Acosta et al., 2013), (Färber et al., 2016), (Kontokostas et al., 2014), (Redman, 2001), (Zaveri et al., 2016) mainly focusing on DBpedia and RDF triples (researching the Semantic Web or converting data to RDF). These researches are useful checking linked (open) data quality but aren't appropriate checking standalone data set quality by non-IT and/or non-DQ expert as these approaches requires deep knowledges in appropriate field.

Another interesting approach uses crowdsourcing handling linked (open) data quality problems that are challenging to be solved automatically (Acosta et al., 2013). The authors implement a quality assessment methodology for linked open data that leverages the wisdom of the crowds in two ways: (1) a contest targeting an expert crowd of researchers and linked data enthusiasts; (2) paid microtasks published on Amazon Mechanical Turk. The results show that the two styles of crowdsourcing are complementary and that crowdsourcing-enabled quality assessment is a promising and affordable way to enhance the quality of linked open data. Unfortunately, it highly depends on the human factor - on involved people and their knowledges in a given field.

Microsoft also proposes data quality tools. Probably the most widely known data quality tool is Data Quality Services (DQS).

DQS is knowledge-driven data quality tool (SQL Server component) designed for data quality analysis and improvement. According to Laudenschlager et al. (2017), DQS main functional components are:

1. **knowledge bases**. They are used to cleanse data (it is necessary to have knowledge about the data). To prepare knowledge for a data quality project, knowledge base, which DQS can use identifying incorrect or invalid data, must be built and maintained;
2. **data matching process**. It enables data duplication reduction and data accuracy improvement. It analyses the degree of duplication in all records of a single data source, returning weighted probabilities of a match between each set of records compared. User can then decide which records are matches and take the appropriate action on the source data.
3. **data cleansing**. It is the process of analysing the quality of data in a data source, manually approving/rejecting the suggestions by the system, and thereby making changes to the data:
   - **the computer-assisted process** uses the knowledge in a DQS knowledge base to automatically process the data, and suggest replacements/corrections. The data processed by DQS is split into 5 groups: suggested, new, invalid, corrected, correct and is available to the user for further processing of the data;
   - **the interactive process** allows the data steward to approve, reject, or modify the changes manually proposed by the DQS during the computer-assisted cleansing.
4. **data profiling** has two major goals: (1) to guide user through data quality processes and support users' decisions, and (2) to assess the effectiveness of the processes. It provides users with automated measurements of data quality.

Although some of DQS functions are very effective and could be used in data analysis, DQS has several disadvantages: (1) it is possible to analyse only one table per time – multiple table analysis isn't available; (2) there exists domain formats limitations; (3) minimal data matching threshold is 80% and it can't be reduced; (4) analysis of bigger data amounts (for instance, few million records) requires high resources (CPU, Disk and Memory usage grows up to 100%).

To sum up, data quality is complex concept which depends on the specific use-case – on data application. Most of the existing researches can't be used by non-IT or/and non- data quality experts as they require deep knowledges not only in IT field but also in data quality field, especially if it is used one of following previously discussed approaches: (Caro et al., 2007), (Ferney et al., 2017), (Neumaier, 2015), (Umbrich et al., 2014), (Vetrò et al., 2016). These approaches can be used by people with an appropriate background in data quality field or involving them in all stages of data quality analysis process as these approaches: (1) use high number of dimensions and it may seem an overstatement, especially for practitioners, and/or (2) require defining data quality requirements, and/or (3) require relating defined or already existing data quality requirements to appropriate dimensions, which would be further applied on the data set. Some of these approaches require knowledges also in linked data, knowledge engineering or specific methodologies.

Specific knowledges, which are necessary to use these approaches, let divide users who will use them in two groups: IT-experts and data quality experts (DQ-experts). Overlapping of these two groups is possible. In previous papers ((Bicevska et al., 2017), (Bicevskis et al., 2018a, 2018b), (Nikiforova, 2018)) only "IT-expert" or "IT-professional" concepts were used. Person was considered as IT-expert if he/she has IT-background. However, this concept isn't precise as not all IT-experts have proper knowledges in data quality field and could be considered as data quality experts. Moreover, it is possible to have appropriate knowledges in data quality field without deep knowledges (such as education or experience) in IT field. Such people can be often met in banks and other fields, for example, data analytics can have proper knowledges to analyse data quality of specific data sets without relation to IT field (just basic knowledges to work with proper technologies and devices). Therefore, in this paper new more appropriate concept is involved - "data quality expert". In this paper person is considered as DQ-expert if it has deep knowledges of concepts, which are related to data quality, and is able to make previously mentioned tasks.

For approaches which requires deep IT knowledges (for instance, (Acosta et al., 2013), (Färber et al., 2016), (Gandhi, 2016), (Kontokostas et al., 2014), (Redman, 2001), (Zaveri et al., 2016)), previously used IT-expert concept is valid. Person is an IT-expert if it has IT-background - education and/or experience in IT field which covers needed topics (specific technologies, approaches, data quality etc.).

One common idea of analysed researches is that (open) data has data quality issues ((Acosta et al., 2013), (Färber et al., 2016), (Ferney et al., 2017), (Gandhi, 2016), (Guha-Sapir and Below, 2002), (Kontokostas et al., 2014), (Vetrò et al., 2016)). It is important to explore open data and offer approaches which could improve open data quality.

**Problem Statement**

Importance of (open) data quality topic was already discussed in Sections 1 and 2. To sum up, open data quality is popular and on-the-run topic due the fact, that open data became popular and widely used but, unfortunately, it may have data quality problems. According to the literature, most of analysed open data sets (Section 2) have data quality issues, and this is common problem for many countries.

Nowadays open data is extremely popular and there are a lot of open data portals in many countries, which provide users with a wide range of different data such as company registers, education and culture, civilians, development, environment and other. Latvia also has such portals - (WEB, h), (WEB, b). (WEB, b) collects 139 different open data sets, which are provided by 41 publishers. Moreover, it is supposed that in the nearest future this list will be supplied with new data sets (for instance, there is planned to publish 8 new geospatial data sets, but in the first two summer months (of 2018) 27 new data sets have been published (WEB, c) etc.). According to (WEB, c), one of the new directions of "Directive on the re-use of public sector information" is opening state enterprise data focusing in high-quality data opening.

According to Global Open Data Index (WEB, o), which provides the most comprehensive snapshot available of the state of open government data publication, Latvia takes the $14^{th}$ place among 94 countries. According to this indexation, Latvia has improved the state of open data quality as in 2014 it took the $24^{th}$ place but in 2015 – the $31^{st}$. It means that this topic is also popular for Latvia, as the number of open data sets increase and data consumers (and suppliers) has necessity to check its quality.

But before open data quality will be discussed in more details, it is important to understand "open data" concept. Open data key idea is to collect data, which will be free available to everyone to use, analyse and republish as they need without any restrictions.

Open data is usually divided in three groups (by its type and source): (a) government data, (b) research or science data and (c) private sector data (Gruen et al., 2014). Government data (or Public-Sector Information) includes a wide range of data collected or funded by national, regional and local governments and government agencies that can be purposefully collected (e.g. national statistics, meteorological, mapping and other spatial data) or arise as an integral part of the government function (e.g. business registration, court records) (Gruen et al., 2014). Research or science data is open data that arises from research that is publicly funded. Private sector data includes such data as vehicle tracking information for traffic management and infrastructure design and development, barcode sales data for economic management, such as estimation of consumer price index. This open data type could be open for both public and private benefit (Gruen et al., 2014).

According to 8 principles, which must be satisfied in order to consider data as open data (Bauer and Kaltenböck, 2011), (WEB, h), (WEB, i), data must be:
1. **complete** - all public data is made available. Datasets released by the government should be as complete as possible, reflecting the entirety of what is recorded about a particular subject. All raw information from a dataset should be released to the public, except to the extent necessary to comply with federal law regarding the release of personally identifiable information. Metadata that defines and explains the raw data should be included as well, along with formulas and explanations for how derived data was calculated. Doing so will permit users to understand the scope of information available and examine each data item at the greatest possible level of detail.
2. **primary** - data is published as collected at the source, with the finest possible level of granularity, and not in aggregate or modified forms.
3. **timely** - data is made available as quickly as necessary to preserve the value of the data. Whenever feasible, information collected by the government should be released as quickly as it is gathered and collected. Priority should be given to data, which utility is time sensitive. Real-time information updates would maximize the utility the public can obtain from this information.
4. **accessible** - data is available to the widest range of users for the widest range of purposes.
5. **machine-processable** - data is structured so that it can be processed in an automated way. For instance, information shared in PDF format, is very difficult for machines to parse, so, this format is not the most appropriate. Thus, information should be stored in widely-used file formats that easily lend themselves to machine processing. These files should be accompanied by documentation related to the format and how to use it in relation to the data.
6. **non-discriminatory** - data is available to anyone, with no registration requirement.
7. **non-proprietary** - data is available in a format over which no entity has exclusive control.
8. **licence-free** - data is not subject to any copyright, patent, trademark or trade secrets regulation. Reasonable privacy, security and privilege restrictions may be allowed as governed by other statutes. Compliance to these principles must be reviewable through the following means: a contact person must be designated to respond to (a) people trying to use the data or (b) complaints about violations of the principles (Bauer and Kaltenböck, 2011), (WEB, h).

Sunlight foundation supplied this list with two more principles (WEB, i):
9. **permanence** which refers to the capability of finding information over time;
10. **usage costs** - one of the greatest barriers to access to ostensibly publicly-available information is the cost imposed on the public for access – even when the cost is *de minimus*.

These principles don't focus on data quality. It let make assumption that open data may have quality defects. According to (Zuiderwijk et al., 2012), data quality is one of the main barriers of open data adoption. Usually open data is used without additional activities (such as quality checks) making assumption that it has been already checked by data suppliers and is ready for further processing. Due the fact this data is used in business decision-making, it is important to be sure that data is of high quality and is trustable.

Author of this paper suppose that (1) open data often isn't checked before it is published and (2) its analysis will detect quality problems. This paper summarizes data quality analysis of Latvian open data sets.

Given analysis is made using approach which is briefly described in the next section (more detailed in (Bicevska et al., 2017) and (Bicevskis et al., 2018a)). This approach is considered as one of the most appropriate to the specified task for several reasons. Firstly, as stated in the previous section, most of the existing solutions are unsuitable for professionals without appropriate IT or DQ background. Moreover, most of these researches can't be applied to the specific data sets, defining data quality requirements for specific columns. Wang and Strong (1996) believed that data consumers have a much broader data quality conceptualization than IS professionals realize. However, definition of data quality dimensions and methods for its quantitative evaluation is one of the most important steps was ever made. Inspired from Wang and Strong (1996), Bicevska et al. (2017) (author's colleagues) proposed mechanisms for data quality characteristics specification in formalized languages, which could provide users with possibility to easily check specific data sets quality, defining specific requirements for specific columns. In this research (Bicevska et al., 2017) approach will be used, verifying its appropriateness to the specified task. Main points of this approach will be briefly discussed in Section 4.

## Research Method

According to TDQM methodology (WEB, j), data quality lifecycle can be described by four interconnected data quality control phases:
1. **data quality definition** - requires defining data quality description and data quality metrics. In scope of this research first phase means data object definition, which quality will be analysed, and quality specification definition which will be used in quality analysis;
2. **data quality measuring** - requires data selection from various data sources and measuring previously defined quality requirements fulfilment, and, when appeared data object fields values irrelevance, corresponding data objects are written to error data protocol;
3. **data quality analysis** at which data quality test results must be analysed with aim to detect data quality problems – it is necessary to find out main reasons of these problems and to choose the most appropriate quality improvement activities;
4. **data quality improvement**.

This research deals with phases 1-3. Data quality improvement phase isn't researched as it can be ensured using existing tools like previously mentioned MS DQS.

Due all phases of described cycle must be systematically repeated to achieve and to keep high data quality, it is important to understand that data quality control process is given as a cycle. First of all, it is necessary to check data quality mechanism efficiency by repetition of the $2^{nd}$ – measuring and the $3^{rd}$ – analysis phases. Secondly, data in data storages is continuously changing and new data can bring new data quality problems and/or cause new requirements. Further, assessment criteria should be changed and new data quality requirements or different data quality conditions could be defined at the new cycle iteration.

Traditional quality checks implementation is hard to test and to manage. It is one of the reasons why data quality is rarely checked, and data often have data quality defects. To solve this problem, they are substituted with universal decision – quality checks are separated from program code. As data quality has relative and dynamic nature - data quality requirements depend on data application and staged data accruing. According to these facts, used approach (Bicevska et al., 2017) ensures possibility to change and define new quality requirements, and its main principles are:

(1) data quality requirements can be formulated on several levels – (a) for specific data object, (b) data object in the scope of its attributes, (c) data object in the scope of database and (d) data object in the scope of many databases;
(2) at the first stage, data quality definition language is graphical DSL (Domain Specific Language), which syntax and semantics can be easily applied to any new information system. It must be simple enough to understand providing industries specialists possibility to define data quality requirements without IT specialists' supervision;
(3) data quality can be checked in various stages of the data processing, each time using its own individual data quality requirements description;
(4) as data quality requirements depend on data application, quality requirements are formulated in platform independent concepts separating quality checks from program code.

Data quality control system of (Bicevska et al., 2017) consists of 3 main components:

1. **data object**: traditionally the notion of a data object is understood as the set of values of the parameters that characterize a real-life object. The research results will be illustrated by simple example from the open data set about Educational Licenses provided by Municipal of Riga. Obviously, data object, which quality will be analysed (see next Section), in this example is Licence (Fig. 1, 2);
2. **quality requirements**: the data quality specification for a specific data object consists of the conditions that must be met in order to consider the data object as of high quality. Firstly, data quality specification may contain informal descriptions of conditions, for example, in natural language or formalized descriptions that are implementation-independent. Processing the data object class, data object class instances are selected from the sources of information and written into collection. Then, all instances are cyclically processed, for each individual instance examining the fulfilment of quality requirements, like the one in the case of processing an individual data object. As a result, the quality problems of each individual instance are detected. Quality requirements are defined in result of analysis of given data set. Each column and data, which it contains, are explored defining the most appropriate data type for given column and quality specifications for data it contains.
   Then, data quality requirements for a data object are defined using logical expressions. The names of data object's attributes/fields serve as operands in the logical expressions.
3. **description of quality measuring process**: the first step in the quality evaluation process describes the activities to be taken selecting data object values from data sources. Thereafter, one or more steps are taken evaluating the quality of the data, each of which describes one test for the compliance of the data object with a specific quality specification. In conclusion, the steps to improve data quality can be followed by automatically or manually triggering changes in the data source. The language describing the quality evaluation process involves verification activities for a particular data object, which can be defined informally as a natural language text, using UML activity diagrams or in the own DSL. Additionally, processing of data object classes instances may require looping constructions. In the example, Fig. 3 shows separate data objects field checks, where each individual operation evaluates the data quality of the field using a SQL statement. The SQL statement SELECT part specifies the target data object, but WHERE - specifies the quality specification. Such a data quality implementation is often found in practice when data is stored in a relational database.

These components form data quality specification, where data object description define data, which quality must be examined. Quality requirements defines conditions must be fulfilled to admit data as qualitative (nullability, format, data type etc.). Description of quality measuring process define procedure which must be performed to assess data quality. Obviously, in the centre of data quality control system is data object. It is a set of existing object parameters. It has only those data, which quality must be analysed. It reduces amount of processed data (kind of denormalization). According to the structure of the data objects class, data object has random number of fields, and their attributes prescribe possible value constraints and other data object classes. Data objects classes are necessary to define data quality requirements for collections of data objects, for example, for database, which stores data about people, invalid person data amount can't exceed 1% of all records - such quality dimension is measurable only if all parameters values of specific record are checked one by one by, relating number of errors to the number of processed records (Nikiforova, 2018).

It is important to create data quality model, which consists of graphical models, where each diagram describe specific part of data quality check. All checks of one business process step are unified in the packages. All packages form data quality model. Diagrams consists of vertexes, which have data quality control activities and are connected with arcs, which specify order of these activities. It can be used in two ways: (1) informal, which has description of necessary checks activities in natural language, where diagram symbols have textual description of activities and (2) executable, which can be get by conversion of informal model substituting informal texts by code, SQL queries or another executable object.

In this research, all these components are described with language metamodels - each component is provided with its graphical representation. 3 language families are defined. Their syntax is defined by representational models supported by tools developing platform DIMOD, which allow to define a variety of DSL with different data objects structures and after its inserting in repository, DIMOD behaves as DSL graphic editor. Involved languages are developed as graphic languages and are related with tools possibilities of the development platform DIMOD, which is advised to be used instead of using specific data objects definition language (Bicevska et al., 2017).

To sum up, one of the key principles of used approach is focusing on data objects (avoiding linking data quality issues by dimension).

Data object definition and data quality requirements specification can be done even by one person. However, more than one person can be involved. In general case, used approach supposes and even support interaction between business-level and technical units (also between different persons or even teams within an organization or many organizations if needed). Definition of data object and data quality specification using diagrams makes interaction process easier – diagrams are easy to read (to understand), to create and to edit. This approach doesn't have specific limitations relating to the teams making data quality analysis - people involved in this process could be from different organizations and can be not belonging to any of them.

This approach captures the aspects of data quality which are important to data consumer.

## Open Data analysis

This section demonstrates open data quality analysis using previously described approach. This paper demonstrates the analysis of 4 Latvian data sets provided by Municipal of Riga – statistics on communication with Riga municipality (WEB, d) and educational licences for 3 years (WEB, h). It also summarizes results of analysis of 2 data sets provided by government – data about government information systems (WEB, e) which is provided by Ministry of Environmental Protection and Regional Development of the Republic of Latvia and The Register of Enterprises of the Republic of Latvia (WEB, m).

First of all, as these portals are main open data portals in Latvia, it is assumed they should have data of higher data quality. Moreover, according to (WEB, g), The Register of Enterprises of the Republic of Latvia (WEB, m) data set has the highest score among Latvian open data sets (90 of 100 points). It is also interesting to analyse one of the potentially best Latvian data sets, which should have the least number of data quality problems (possibly, it won't have any data quality problems). However, according to Global Open Data Index (WEB, o), The Register of Enterprises of the Republic of Latvia takes the 18$^{th}$ place among 94 analysed Company registers.

Secondly, as analysed data sets are from different data suppliers, it is possible to make more general conclusions on data quality of Latvian open data. Data sets from different data suppliers allow checking if there exist differences between data quality of data provided by governments (the centralized data release, at national level) and municipalities (decentralized data release). This assumption is made according to (Vetrò et al., 2016)).

Analysed data sets satisfy the most part of open data basic principles (see Section 2). Unfortunately, these data sets aren't accompanied by documentation or description related to the format and how to use them in relation to the data. However, (WEB, d) and (WEB, m) have short textual description of data sets fields (without data format requirements). Timeliness of most part of these data sets is hard to check, however, some of them, for instance, The Register of Enterprises of the Republic of Latvia and Statistics on communication with Riga municipality, are updated once a day.

Municipal of Riga provides data about educational licenses for 2013, 2014, 2015. Data is available in two formats - .xls (Microsoft Excel) and .csv (Comma Separated values) and consists of 9 columns: license requester, registration number, realized program, program type, realization place, hours, decision, terms, license number. Initially, all data format is varchar but in the result of data exploration (analysing each column), more appropriate data formats, nullability constraints and other data quality requirements for each field were defined. Defined data quality requirements are data consumers interpretation of data. Obviously, if data suppliers make data analysis, more accurate data quality requirements could be defined. Nullability constraints were defined relating NULL values of specific column to the total number of records and if this rate is lower than 3%, it is considered specific field can't store NULL values (excepting cases where data must be provided, for instance, primary data for specific data object). As analysed data sets have the same structure, Figures 1-3 can be applied to all of them (making appropriate changes).

Figure 1 describes data in its original state (input data) in informal way.

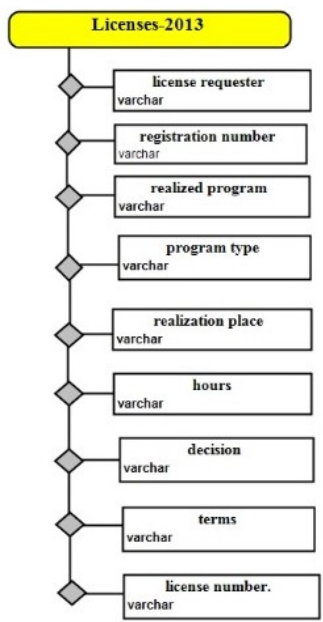 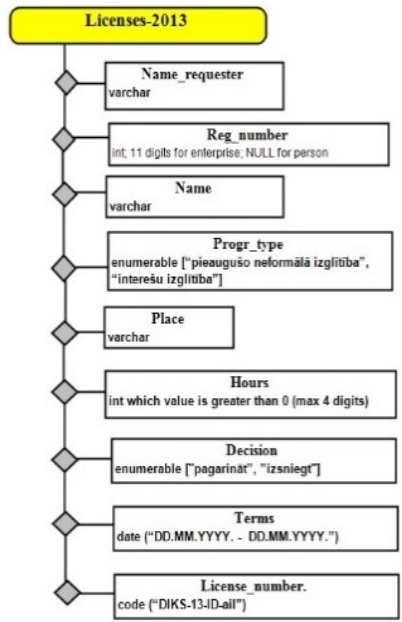

Fig. 1. Data object "Licences-2013" (input)    Fig. 2. Data object (database)

Figure 2 shows data state in database with value attributes and data quality specification, which is used in data quality requirements specification.

When data quality requirements are inspected, data object fields attributes can be changed to attributes, which are shown on Figure 3. At this stage of the research, analysed data sets were imported to SQL Server database and all records were processed with SQL queries according to defined quality requirements and developed diagrams. In the future, this process will be automatized making diagrams executable.

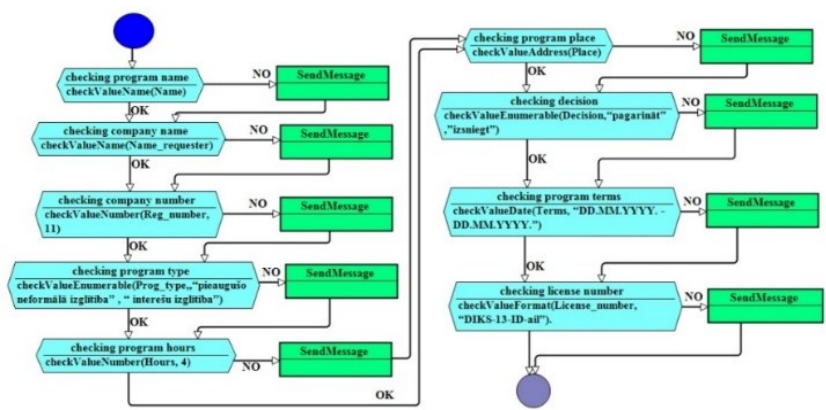

Fig. 3. Data object "Licences-2013" quality specification

The 4th data set provided by Municipal of Riga - Statistics on Communication with Riga Municipality is available in .csv format. It has 39 490 records and has 8 columns, which stores record ID number, communication direction, communication channel, topic group and topic, client type, record date and number of records ("count"). As data object and quality specification are defined according to previously described principles, corresponding figures are not provided in this paper.

The 5th data set - Data about Government Information Systems is available in .csv format. This data set summarizes data about government information systems, their controllers and higher authorities. Each data object - government information system is described by 36 parameters (columns) such as name, address, number, accruing of personal and financial data, data transmission protocols, users, carriers' data, responsible person data, website etc. Data quality requirements for some of these fields are available in Table 1 (Section 6) together with results of this data set analysis and detected quality problems.

The 6th data set - The Register of Enterprises of the Republic of Latvia has 22 columns - 396 952 records (proposed ideas can be also applied to larger data sets). Data is available in two formats - .xls and .csv.

3 more Company Registers of other European countries (United Kingdom, Estonia, and Norway) are briefly discussed in the next Section. It should be also mentioned, that according to Global Open Data Index (WEB, p), Company register of the United Kingdom and Norway takes the 1st place (while The Register of Enterprises of the Republic of Latvia is lower – at the 18th place). Despite this Index focuses on evaluation of data sets openness, author suppose that these data sets could have less data quality problems.

The results of data analysis are available in the next section.

## Results

This paper covers 9 data sets analysis. These data sets were imported to SQL Server database and processed according to the proposed approach – defining data object, which quality should be assessed, and data quality requirements, which data must fulfil (records were processed with SQL queries according to defined quality requirements). Each component is provided with its graphical representation.

The 1st – 3rd processed data sets (Educational Licences) don't have any significant quality weaknesses. All fields except "stundas" (hours) have values. As "stundas" (hours) field was partly filled in all 3 data sets (89% of records have NULL values), author suppose this field can had NULL values. Other fields values, which must be NOT NULL, corresponds to data quality specification. Author concluded that given data sets are of high quality and can be used in analysis and business decisions-making.

However, the 4th data set provided by Municipal of Riga - Statistics on Communication with Riga Municipality has several data quality problems and anomalies:

- According to description of this data set provided by data supplier, only 3 contact channels (field "channel") are valid (3 values are allowed): "e-mail" ("e-pasts"), "portal" ("portāls") and "meeting" ("tikšanās"). However, 6 more values were observed (in total 25.9% of records). Despite the fact all these values make sense (for instance, "phone", "short message", "other", "fax") they aren't supposed to be there as description contains only 3 values. It means that there consist inconsistencies between data set description and accrued data.
- "Topicgroup" field has 26 different values but according to analysis, there are 4 different names which occurs only once, 3 – 2 times and 1 – 3 times (0.02% records) – other 23 values appears more often. These values could be considered as data anomalies, which should be checked with data suppliers.

In total, 2 of 8 columns (25%) of the 4th data set - Statistics on Communication with Riga Municipality have data quality problems (insignificant data quality problems or anomalies) which should be revised by data supplier solving them (the first problem possibly could be solved adding 6 more values to the list of contact channels). Other fields correspond with defined quality requirements and don't have any quality problems.

Error rate is calculated by relating number of detected errors to the total number of processed records.

Overall data quality of data sets provided by Municipal of Riga can be considered as high. These data sets can be used in analysis and decisions-making. However, the 4th data set has some defects. Detected data quality problems and anomalies could be easily solved as their number is quite low. Correction of these problems doesn't require high resources.

In case of Data about Government Information Systems (the 5th data set), 25 of 36 columns (69.4%) have at least few (in some cases insignificant) data quality problems (see Table 1 for data quality requirements and overview on detected quality problems):

- The most common data quality problem of this data set is different values pointing to the absence of data - '-', 'no' ('nav') values instead of (or together with) NULL values.
  One possible explanation could be design of relational database, which doesn't allow to leave particular fields blank, however, 12 fields (33.3% of columns) have previously mentioned values despite NULL values are also met in these fields. From the one hand, such data quality problem could be considered as insignificant, but as data quality depends on the use-case, in specific cases it can be important and could affect analysis results. Simple example of importance of this data homogeneity is the case, when it is necessary to count number of existing records (has value) and

- non-existing values (field is blank - doesn't meet specific condition, for instance, objects that don't have web page or can't be contacted via post). In general, such analysis could be easily done analysing blank fields (where specific field has NULL value – is not filled) but in this case, even filled fields can store value meaning data absence ('-', 'none') and if user don't know it, analysis results will be inaccurate - data can't be used in analysis.
- Closing date should be fulfilled only if the system is closed ("Active or closed system" ("Aktīvā vai slēgtā sistēma") field value is "closed" ("slēgts")). According to this requirement, closing date is NULL only if the previous field value is NULL – this requirement is fulfilled. Moreover, all stored dates are valid, however, 11% of records have date format "MM.DD.YYY" and doesn't correspond to quality requirements – "MM/DD/YYYY". This quality problem is insignificant as all values are valid (format "mdy" is used in both cases) and the only difference is used separator.
- 20% of records has quality problems in "website" ("Tīmekļvietne") field as values don't correspond to website address pattern – aren't valid addresses. 35.14% of information systems don't have website. However, only 11.43% of records has NOT NULL values, which points on address inexistence. 4 different values are used to point address doesn't exist: 'http://-' (9%), 'http://' (0.82%), 'http://Nav' (0.4%) and 'http://Nav%' (1.22%).
- 0.82% records don't have responsible person phone number and it means that these information systems can't be contacted via phone. Moreover, 1.22% of information systems have also invalid e-mail address. However, responsible person name, surname, company and company address is filled for all objects.
- 5.71% of records have invalid carriers and 1.63% - managers authority codes while 0.82% of carriers and managers codes are empty and 0.82% of managers codes has '-' values.

Correction of detected data quality problems won't require a lot of work as part of problems of missing values is related to one record where 17 of 36 fields were blank and this records correction would improve mentioned results.

**Table 1.** "Data about Government Information Systems". DQ requirements and analysis results.

| Field name | Format, Quality requirement | Error rate (%) | Comment, problem |
|---|---|---|---|
| IS number | Int, NOT NULL | 0 | - |
| IS name | Varchar, NOT NULL | 0 | - |
| IS have personal data | Varchar, 2 allowed values: "contains"/ "doesn't contain", NOT NULL | 0.4% | NULL values |
| IS have financial data | Varchar, 2 allowed values: "contains"/ "doesn't contain", NOT NULL | 0.4% | NULL values |
| Service rate | Varchar NOT NULL | at least 14.28% at most 32.68% | 0.4% – NULL value 18.4% '-' values 13.88% 'no' ('nav') values |
| Website | Varchar match pattern "http://www.%.%" or "http://%.lv%" NULL | At least 20% | 20% invalid addresses (don't match pattern); 23.7% NULL values; 11.4% (4 different values) point on address inexistence: "http://-"– 9%; "http://" – 0.82%; "http://Nav" – 0.4%; "http://Nav%"" – 1.22% |
| Responsible person number | Int NOT NULL | 0 | - |
| Responsible person name | Varchar, NOT NULL | 0 | - |
| Responsible person phone number | Int, (8 digits length, starts with '6') or (11 digits length, starts with "371"), NOT NULL | 0.82% | incorrect format |

| Officer e-mail | Varchar, match pattern "%@%.lv", NOT NULL | 1.22% | invalid addresses |
|---|---|---|---|
| Manager code | Int, 11 digits, NOT NULL | 1.63% | 0.82% NULL values<br>0.82% incorrect formats |
| Holder code | Int, 11 digits, NOT NULL | 5.71% | 0.82% NULL values<br>0.82% '-' values<br>4.1% incorrect formats |
| Holder name | Varchar, NOT NULL | 2% | '-' value |

Results of analysis of The Register of Enterprises of the Republic of Latvia are very unexpectable – it has many quality problems (see Table 2 for overview on quality requirements and detected quality problems). 11 of 22 fields – 50% have data quality defects. For instance:

- 10 enterprises don't have name (error rate is 0.0025%);
- 94 records (0.024%) don't have registration date;
- 366 records (0.09%) don't have "Address" and 4 523 records don't have "Addressid" (1.14%). Moreover, 2 records have "Adressid" value but don't have address even though these fields are connected and if one value exists, the second must exist, too – both fields must be filled (number of empty values also should be equal).

However, data format of all NOT NULL values corresponds to data quality specification.

- 1 403 (0.35%) records don't have "Type_text" despite field "Type", which has an abbreviation of "Type_text", is NOT NULL. 6 forms of organization - ASF, KOR, PRO, SAA, SPA un SPO don't have "Type_text" value (full-text for an abbreviation);
- 20 496 (5.16%) records don't have index and 2 indexes have incorrect format - 3 digits instead of 4;
- 1.15% of records don't have ATV code (administrative-territorial index) and 0.24% values format is incorrect - have less than 7-digit;
- 646 (0.18%) enterprises have "terminated" date but don't have "closed" value – "terminated" date can have only those enterprises, which are already closed – liquidated or reorganized.
- 94 records don't have registration date (error rate - 0.024%). "Registered" field can't have NULL values.

**Table 2.** "The Register of Enterprises of the Republic of Latvia". DQ requirements and analysis results.

| Field Name | Format,<br>Quality requirement | Error Count | Error Rate (%) | Comment, Problem |
|---|---|---|---|---|
| Reg_number<br>(registration number) | Int, 11 digits<br>NOT NULL | 0 | 0 | - |
| Name | Varchar, NOT NULL | 10 | 0,0025% | NULL values |
| Type<br>(company form - abbreviation) | Varchar,<br>Enumerable {Ltd, etc.} | 0 | 0 | - |
| Type_text<br>(form – full-text) | Varchar, NOT NULL | 1 403 | 0.35% | NULL values |
| Address | Varchar, NOT NULL | 366 | 0.09% | NULL values |
| Address_id | Int, 9 digits, NOT NULL | 4 523 | 1.14% | NULL values |
| Region_code | Int, 9 digits, NOT NULL | 280 662 | 70.7% | NULL values |
| City-code | Int, 9 digits, NOT NULL | 99 049 | 24.95% | NULL values |
| Post-code | Int, 4 digits,<br>NOT NULL | 20 498 | 5.16% | 2 – incorrect format (3 digits instead of 4);<br>20 496 NULL values |
| ATV-code<br>(administrative-territorial index) | Int, 7 digits,<br>NOT NULL | 5 521 | 1.39% | 4 574 NULL values;<br>947 – incorrect format (less than 7 digits); |

Same approach was applied to another 3 Company Registers of other European countries (United Kingdom, Estonia and Norway). Analysis results of these data sets demonstrate that all data sets have at least few quality problems:

- Company Register of United Kingdom - in 15 of 55 columns (27.3%),
- Norwegian register – in 8 of 42 columns (19%),
- Estonian register – in 7 of 14 columns (50%).

The most common data quality problems are similar with data quality problems detected in the previously mentioned Register of Enterprises of the Republic of Latvia. Most of the quality problems are detected in (1) address fields - NULL values and (2) date fields - dubious date values, for example, according to the registers, there are companies which were organised in 1552, (3) fields storing abbreviation – NULL values while text field with explanation for this specific abbreviation is provided and versa. There are also some specific data quality problems such as: (1) fields which are supposed to store names of countries store postal codes or name of counties, or name of specific area (2) various names for one country in one data set (UK and United Kingdom; USA, United States and United States of America etc.) or name of disbanded countries (despite the fact company was registered after specific country was disbanded); (3) codes doesn't correspond with the specified format or length; (4) specific fields or few records are filled with 'x' char or '0'. More detailed analysis of these data sets with description of more specific data quality problems for each analysed Company Register, is available in (Bicevskis et al., 2018b). These results show that processed data set has quality problems even in those fields, which have primary company information – name, registration number and registration date. It means that not all companies could be identified and contacted (as address field has NULL values). Some of these records has inconsistent data (active companies have date of termination). Moreover, there is no confidence that these companies exist (possibly, these records were added as test data or they are too old but nobody has deleted them). Analysis results allow to conclude that only Estonian and Norwegian company registers could be used identifying all registered companies (by their name, registration number and incorporation date) as another 2 Company Registers have few data quality problems in these fields. It should be also mentioned that Estonian Company Register doesn't provide incorporation dates (company identification is possible by name and registration number) but Norwegian Company Register have 9 records with invalid liquidation date. Moreover, none of these data sets could be used contacting every registered company via post (using address and postal code). It does not mean that the data from company registers cannot be trusted when a company must be identified since the number of defects is insignificant. Detected data quality problems could be easily solved using given approach identifying them. It would significantly improve overall data quality of these data sets.

These results correspond to Global Open Data Index (WEB, p), according to which Company register of the United Kingdom and Norway rank is better than The Register of Enterprises of the Republic of Latvia. However, all analysed data sets have data quality problems and, moreover, Company register of the United Kingdom, which, according to Global Open Data Index, takes the 1st place, has data quality problems even in fields storing primary data about companies. It means that authors' assumption that data sets, which takes the highest place, won't have data quality problems, was incorrect.

Despite the fact the 5th and 6th data sets have a high number of data quality problems, it is not correctly to admit them as data sets of low data quality as data quality depends on the use-case. It means that, if the specific data set has lot data quality problems, it still can be used and for some purposes it could appear as the data set of high quality, if the user is interested in fields storing data without quality problems. For instance, let's assume the 5th data set - Data about Government Information Systems - is used to get data about information system – its name and registration number, status and responsible person – its name, surname and registration number. In this case provided data is of high quality and could be used making analysis as these fields don't have any data quality problems. It is just one example when this data set can be used getting trustable results. At the same time, let's assume, we want (1) to get summary on systems, which do or doesn't store personal or financial data, based on appropriate field with two possible values ("contains"/ "doesn't contain") or/and (2) in addition to information system name and registration number we want to get its website address if it exists or/and (3) we need to contact responsible person via phone or e-mail, or/and (4) we need to know not only managers or holders name and surname but also their codes. In these cases, this data set is not qualitative enough as some data is missed and some data is incorrect, invalid. Unfortunately, users even don't suppose that some fields have data quality problems and especially which records are defective. In these cases, this data set must be processed before it will be used in analysis.

In this research, given analysis checks (1) existence of values, (2) relevance to the specified data type, (3) format of stored values (for example, length of the stored value), (4) conformity to the specific pattern, (5) relevance to the list of enumerable values, (6) validity of value (for example, trustful date) and other conditions which arise from the data sets and type of value which could be stored in the specific fields.

Main results of these analysis are that (1) proposed approach lets analyse data quality of data sets without knowledges about how this data was collected and processed by data suppliers – it is possible to analyse

"foreign" data, (2) detect quality problems and anomalies in open data, which is available for every stakeholder and can be used in business decision-making, (3) open data often has data quality problems, which must be detected before data will be used in analysis, where data quality weaknesses could lead to huge losses.

There also can be concluded that in this specific analysis data sets provided by municipalities sometimes are of higher quality than data sets provided by governments. As it was demonstrated in this section, all data sets provided with municipalities were of high quality – just one of analysed data sets (Statistics on Communication with Riga Municipality) had few data quality problems. At the same time, data sets provided by governments had more data quality problems and in specific cases could be considered as data sets of not high enough quality (depending on the use-case) as all analysed data sets had data quality problems. Moreover, analysis results of data sets provided by governments are valid not only for Latvian data sets (in case of Company Registers) but also for European open data. However, the number of analysed data sets are not high enough to maintain that this tendency will be valid and data quality depends on provider in all cases.

## Future Plans

As was previously stated in (Bicevska et al., 2017), data quality management mechanisms should be able to execute data quality specifications. To check results, which were provided in the previous section, it is planned to achieve previously stated aim – to make quality model executable involving interpreter, which will (1) translate defined diagrams and (2) execute data quality checks, resulting records which have errors. When this step will be completed, it will be possible to analyse open data automatically, using the data quality model for data quality measurement. Currently, analysis was made manually - writing SQL queries and executing them one by one. Analysis results will be the same with results shown in this paper but process of quality checks (1) will be easier and faster (as there won't be necessity to execute SQL queries one by one) and (2) could be used even by non-IT/non-DQ experts providing industries specialists with possibility to define data quality requirements with minimal IT specialists involving.

Another further step is to check contextual interlinks with other data objects - the compliance of data with the true characteristics of a natural data objects (e.g. Company or Licence) as this research covered only the syntactic accuracy.

Results of analysis will be given to data suppliers letting to review them and to solve detected quality problems. Some of detected problems – "anomalies" (see Section 6) could be ignored as defined quality requirements is authors interpretation of stored data. It is also planned to contact representatives of analysed company registers (1) to provide them summary of detected data quality problems (allowing them to approve or decline detected data quality problems) and (2) find out causes of these problems, which will allow to formulate and provide guidelines for other company registers, which should be taken into account accruing and publishing their data. These steps will help achieving higher quality of published open data.

In the future, it is planned to advice proposed and used approach to students letting them evaluate it – applying it to real data sets. This experiment could provide summary on approach and its advantages and/or disadvantages. Moreover, it is planned to compare developed approach with already existing approaches (especially (Caro et al., 2007), (Ferney et al., 2017), (Laudenschlager et al., 2017), (Vetrò et al., 2016)), applying both approaches to the same data sets. This experiment will provide possibility (1) to compare pros and cons of used approach in comparison with already existing, (2) to approve or decline assumption that already existing approaches are unsuitable for non-DQ/non-IT experts but also (3) to compare analysis results – detected data quality problems.

Proposed approach will be applied to more open data sets (not only Latvian) making more general conclusions on open data quality. This approach was already applied to 9 data sets (6 Latvian data sets), however this number is not enough to make general quantitative conclusions on open data. Future analysis will cover more open data sets.

## Conclusions

The paper is a continuation of author' researches in the area of open data quality (Nikiforova, 2018), applying (Bicevska et al., 2017) approach to Latvian open data sets evaluating their quality. Open data analysis results draw the following conclusions:
- Open data is published for many years and this research proves that (Latvian) open data may have data quality problems (even more problems that were supposed) as there are no centralized checking of open data quality and there are not enough researches on open data quality. Stakeholders who

- use data in business decisions-making should know that data sets may have data quality problems and take in to account as it could have huge impact on their business.
- Open data quality depends on data provider but even trusted open data sources (as centralized data releases) may have data quality problems.
- Used approach works and allows (1) to define data objects, which quality will be analysed, retrieving them from different "external" sources and (2) to make data quality assessment detecting data quality problems. As the definition of data object and data quality specification is done using diagrams, interaction process between a wide range of users (e.g. both IT and domain experts or another stakeholders) become easier – diagrams are easy to read, to edit and to understand. Moreover, this approach helps finding anomalies in analysed data. It could be considered as new step in data quality management. It allows to analyse data without any limitations such as format or data amount – it is universal for many purposes. The used approach is not necessarily tailored for open data.
- Open data quality must be inspected because, as it was shown in this research, it can have data quality problems, which must be detected, explored and solved. Proposed approach is one of the most appropriate options that could be used to improve data quality even by non-IT/non-DQ experts.

This research must be continued automatizing proposed approach involving interpreter, which will make data quality model executable but data quality checking process easier and usable even for non-IT/non-DQ experts. This step could lead to global data quality improvements.

## Acknowledgments

This work has been supported by University of Latvia project AAP2016/B032 "Innovative information technologies".